# Sol-gel transition induced by alumina nanoparticles in a model pulmonary surfactant

**Jean-François Berret**[1*], **Fanny Mousseau**[2], **Rémi Le Borgne**[3] et **Evdokia K. Oikonomou**[1]

[1]Université de Paris, CNRS, Matière et systèmes complexes, 75013 Paris, France
[2]Laboratoire d'Optique et Biosciences, École Polytechnique, Institut Polytechnique de Paris, CNRS, INSERM, Route de Saclay, 91128 Palaiseau, France
[3]Université de Paris, CNRS, Institut Jacques Monod, F-75013 Paris, France

**Abstract:** Inhaled airborne particles smaller than 100 nm entering the airways have been shown to deposit in significant amount in the alveolar region of the lungs. The interior of the alveoli is covered with a ~ 1 µm thick lining fluid, called pulmonary surfactant. Inhaled nanoparticles are susceptible to interact with the lung fluid and modify pulmonary functions. Here we evaluate the structural and rheological properties of the pulmonary surfactant substitute Curosurf® which is administered to premature babies for the treatment of respiratory distress syndrome. Curosurf® is considered a reliable model of endogenous pulmonary surfactant in terms of composition, structure and function. Using active microrheology based on magnetically actuated wires, we find that Curosurf® dispersions exhibit a Newtonian behavior at lipid concentration from 0 to 80 g L$^{-1}$, and that the viscosity follows the Krieger-Dougherty law observed for a wide variety of colloids. Upon addition of 40 nm alumina nanoplatelets, a significant change of the Curosurf® rheology is noticed. The dispersions then enter a soft solid phase characterized by an infinite viscosity and a non-zero equilibrium elastic modulus. The sol-gel transition induced by the nanoparticles is interpreted as the result of the alumina/vesicle interaction, which are illustrated by transmission electron microscopy. It also suggests a potential toxicity associated with the modification of the lung fluid structural and dynamical properties.

# I – Introduction

Our respiratory system is made up of channels and ducts by which air enters through the nose and mouth and passes into the airways to reach the lung alveoli during breathing. Of the order of 300 million in adult lungs, the alveoli provide a large area, around 70 m$^2$ that favors the transfer of the oxygen into the bloodstream and the carbon dioxide from the bloodstream into the exhaled air [1, 2]. In humans, the alveoli are 200 µm in size and made of an epithelium wall of thickness around 1 µm. One side of this epithelium is covered with lung lining fluid, also called pulmonary surfactant and on the other side it is surrounded by blood capillaries (**Fig. 1a**) [3]. The pulmonary surfactant is made of phospholipids and proteins in a ratio 90:10, and its physiological concentration has been estimated to be approximately 40 g L$^{-1}$ [4-7]. It is now accepted that the pulmonary surfactant phase represents an active barrier of protection against external particles and pathogens. Depending on their size, the particles may eventually cross the epithelial barrier and enter the bloodstream. The mechanism of passage through the epithelium barrier, known as translocation, is of paramount importance for toxicity and pharmacological studies (**Fig. 1b**) [8]. The translocation processes are complex and, despite continuous efforts and extensive experiments, they are not yet understood.

The deposition profile of airborne particles in the lungs depends on several factors, such as the size, nature and surface reactivity of the particles, as well as the breathing scenario. To provide a quantitative estimation of the regional fractional deposit along the pulmonary track, we used the Multi-Path Particle Dosimetry software (MPPD from Applied Research Associates, Inc.) [9]. The main feature of the regular breathing scenario is that large particles (> 1 µm) are deposited in the upper airways, whereas nanometric ones are able to reach the alveolar region [8, 10, 11]. It has been reported that inhaled particles of sizes less than 100 nm end up significantly deposited in the alveoli, with a regional deposition probability as high as 50% [10]. As such, these nanoparticles are likely to interact with the pulmonary fluid.

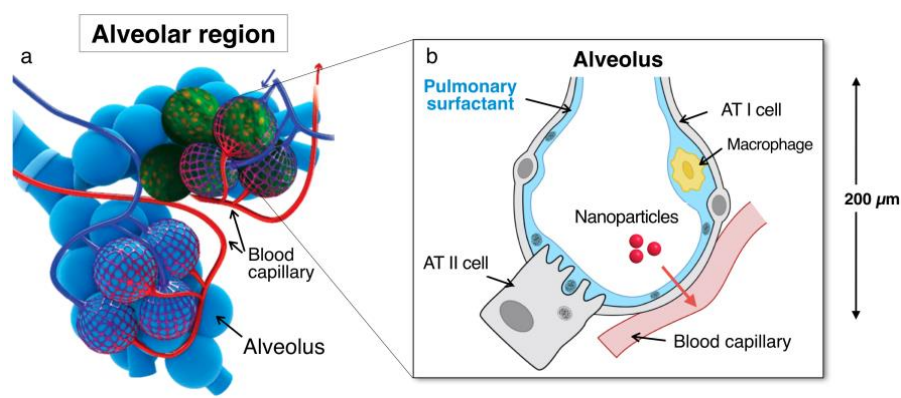

*Figure 1 : a) Schematic representation of the alveolar region. b) Sketch of a human pulmonary alveolus showing the possibility for inhaled particles of nanometric size to cross the air-blood barrier (indicated by the red arrow). In the diagram, AT I and AT II cells refer to alveolar Type I and Type II pneumocytes, the former covering 95% of the internal surface of the alveolus. The pulmonary surfactant lining the alveolar epithelium is produced by the AT II cells. The panel on the left showing the several alveoli is adapted from Ref. [1].*

Recently, we have investigated the interactions of 40 nm organic and inorganic particles with Curosurf®, a lung fluid substitute used as a medication for treatment of respiratory distress syndrome [12, 13]. Curosurf® is given to preterm infants who are at high risk of developing acute respiratory distress syndrome (ARDS), a syndrome associated with lung fluid deficiency and hyaline membrane disease. Randomized studies carried out with Curosurf® have shown a single dose of 200 mg kg$^{-1}$ instilled directly into the child's lower trachea is accompanied of a significant reduction in ARDS severity, with excellent survival rates [14]. In practice, pulmonary surfactant substitutes such as Curosurf® are easier to work with than endogenous surfactant and are considered reliable surfactant models. In particular, they have excellent physico-chemical stability over time, which allows extended characterization of its structure. Curosurf® bulk solutions are dispersions of spherical multivesicular vesicles composed of a mixture of phospholipids and the surface-active proteins SP-B and SP-C [12, 14-18]. Studies have shown that its interaction with nanoparticles is driven by electrostatics, and result in the formation of micron size aggregates, where vesicles maintain their structure and trap the particles at their surfaces. The agglomeration of particles in surfactant phase is a phenomenon of importance as it could modulate the particle interaction with the epithelium, increase uptake and translocation. In a previous paper we focused on the viscosity change induced by various metal oxide particles [19]. Here we focus on the sol-gel transition that occurs in Curosurf® as a function of lipid concentration with the addition of alumina nanoplatelets ($Al_2O_3$-NPs) at the physiological concentration. We also discuss the impact of these transitions on the flow properties of the pulmonary surfactant during respiration.

## II - Materials and Methods

### II.1 – Materials

Curosurf® (Chiesi Pharmaceuticals, Italy) is a porcine minced pulmonary surfactant extract used in neonatal and maternity hospitals mainly in Europe [15]. It is produced as a 80 g $L^{-1}$ suspension containing phospholipids such as phosphatidylcholine (PC) lipids, sphingomyelin (SM), phosphatidylethanolamine (PE), phosphatidylinositol (PI), phosphatidylglycerol (PG) and the hydrophobic proteins SP-B and SP-C [16, 20]. Curosurf® is administered by intratracheal instillation to premature newborns with respiratory distress syndrome. The optimum dose of 200 mg per kilogram body weight is given in a single bolus by highly trained staff. Three milliliters of Curosurf® at 80 g $L^{-1}$ are worth about 1000 €. Curosurf® was kindly provided by Dr. Mostafa Mokhtari and his team from the neonatal service at Hospital Kremlin-Bicêtre, Val-de-Marne, France. The suspension appears as a whitish and low viscosity fluid.

Aluminum oxide ($Al_2O_3$) nanoparticle powder (Disperal®, SASOL, Germany) was dissolved in a nitric acid solution (0.4 wt. % in deionized water) at the concentration of 10 g $L^{-1}$ and sonicated for one hour [13]. In transmission electron microscopy, the particles appear as irregular platelets of diameter 40 nm and thickness 10 nm. In solution (pH 4), their colloidal stability is ensured by electrostatic repulsions mediated by cationic surface charges (charge density +7.3$e$ $nm^{-2}$), resulting in an hydrodynamic diameter $D_H$ = 64 nm [21].

### II.2 – Structural characterization techniques

The hydrodynamic diameter $D_H$ of Curosurf® vesicles was determined on dilute solutions using a NanoZS Zetasizer (Malvern Instruments). The second-order autocorrelation function was recorded in triplicate (T = 25 °C) and analyzed using the cumulant and CONTIN algorithms to determine the average diffusion coefficient, as well as the intensity distribution of the hydrodynamic diameter. Laser Doppler velocimetry using the phase analysis light scattering mode was performed also using the NanoZS Zetasizer to determine the electrophoretic mobility and zeta potential. At physiological pH, the Curosurf® vesicles were found to be negatively charged, with zeta potential values $\zeta$ = -54 ± 8 mV [22]. Nanoparticle Tracking Analysis (NTA) measurements were performed with a NanoSight LM14 (Malvern Instruments, UK), equipped with a sample chamber illuminated by a 532-nm laser. Liquid samples were injected in the measuring chamber until the liquid reached the tip of the nozzle. The software used for recording and analyzing the data was Nanosight NTA 3.0. The Curosurf® concentrations used in the previous experiments have been adjusted so that the results of each technique are optimized. In particular, for NTA and NanoZS studies, concentrations were $10^{-3}$ g $L^{-1}$ and 1 g $L^{-1}$, respectively.

### II.3 – Transmission electron microscopy

For Cryo-TEM, few microliters of a 5 g $L^{-1}$ Curosurf® dispersion were deposited on a lacey carbon coated 200 mesh (Ted Pella Inc.). The concentration of 5 g $L^{-1}$ was chosen because it gives reliable results in terms of the number of observable vesicles. The drop was blotted with a filter paper using a FEI Vitrobot[TM] freeze plunger. The grid was then quenched rapidly in liquid ethane to avoid crystallization and later cooled with liquid nitrogen [23]. The membrane was transferred into the vacuum column of a JEOL 1400 TEM microscope (120 kV). Note that the cryoTEM technique has also limitations, particularly with respect to the maximum size of objects that can be visualized. For deformable colloids, this size is estimated to be 1 µm. It was hypothesized that vesicles larger than 1 µm are expelled from the water film during blotting [24].

For TEM, a dispersion was prepared by simple mixing Curosurf® and $Al_2O_3$-NPs, their final concentrations being 80 g $L^{-1}$ and 1 g $L^{-1}$ respectively. The dispersion was fixed with 2.5% glutaraldehyde and 4% PFA in PBS buffer. The sample was stored in the fixative for 3 h at room temperature and then kept at 4 °C. Afterwards, the sample was centrifuged at 500 – 1000 g for 3 min to obtain a pellet. After several washing steps in buffer, the sample was subsequently post-fixed in osmium tetroxide, dehydrated in an ascending ethanol series and embedded in Epon at 60°C. The Eppendorf cups were then removed, and ultrathin 70-nm sections were cut using an ultramicrotome UC6 (Leica). The sections were analyzed with a 120-kV transmission electron microscope Tecnai 12 (ThermoFisher Scientific) using the 4-k camera OneView and the GMS3 software (Gatan).

**II.4 – Phase-contrast and bright field optical microscopy**
Phase-contrast and bright field images were acquired on an IX73 inverted microscope (Olympus) equipped with 20× and 100× objectives. An EXi Blue camera (QImaging) and Metaview software (Universal Imaging Inc.) were used as acquisition system.

**II.5 – Rheology**
Rheology experiments were performed using a Physica RheoCompass MCR 302 (Anton Paar) working with a cone-and-plate geometry (diameter 50 mm, cone angle 1°, sample volume 0.7 mL). The MCR 302 rheometer was operated in rate-controlled mode for the measurements of the storage and loss moduli $G'(\omega)$ and $G''(\omega)$ and of the stress *versus* shear rate curves $\sigma(\dot{\gamma})$. The macrorheology was carried out in duplicate at T = 25 °C and 37 °C [25]. Pertaining to their frequency and shear rate dependences, the data for elastic moduli and for the shear stress were similar at the two temperatures tested, corroborating earlier results from King *et al.* on other surfactant mimetics (Infasurf) [26]. The magnetic wire microrheology technique has been described in previous accounts [27-29]. Curosurf® stock suspensions were used as received and diluted to the desired concentrations of 5, 20, 40, 50, 70 and 80 g $L^{-1}$ using PBS. A total of $10^5$ wires (contained in 0.5 µL) was then added to 100 µL of the previous suspensions and gently stirred. 25 µL of the previous suspension were deposited on a glass plate and sealed into to a Gene Frame® (Abgene/Advanced Biotech, dimensions 10×10×0.25 $mm^3$). The microrheology protocol is based on the magnetic rotational spectroscopy (MRS) technique. MRS consists in applying a rotating magnetic field to a wire and recording its motion by time-lapse microscopy. For calibration, MRS was performed on a series of water-glycerol mixtures of increasing viscosities.

# III – Results and discussion
**III.1 – Structure of Curosurf® dispersions**

**Fig. 2a** shows an image of a Curosurf® dispersion obtained by Cryo-TEM on a 5 g L$^{-1}$ sample. The sample was obtained by diluting Curosurf® clinical formulation at 80 g L$^{-1}$ with DI-water. The Cryo-TEM image displays multi-lamellar and multi-vesicular vesicles ranging in size from 50 to 500 nm. According to the phospholipid terminology, multi-lamellar vesicles exhibit an onion-like layered microstructure whereas multi-vesicular vesicles are compartments encapsulating smaller vesicles without concentric arrangement. The Cryo-TEM results of **Fig. 2a** are in good agreement with those of the literature [30, 31], as well as with the data on porcine surfactant [32]. Additional studies using the nanoparticle tracking analysis (NTA) technique on dispersions diluted in PBS ($c = 10^{-3}$ g/L) have allowed to determine the number distribution of vesicles as a function of their size. **Fig. 2b** shows that this distribution has a broad maximum around 150-200 nm, followed by a shoulder around 400 nm. This technique allows calculation of the volume fraction of vesicles for a given concentration. Here we found a vesicle density of 7.4×10$^{11}$ L$^{-1}$, leading to the relationship $\phi = 5.2 \times 10^{-3} c$, where $c$ is expressed here in g L$^{-1}$. In the sequel of the paper, we assume that the $\phi(c)$-relation is verified at higher Curosurf® concentration as well. As a result, the concentration at which Curosurf® is formulated (80 g L$^{-1}$) is associated with the volume fraction of $\phi$ = 0.42, suggesting that the vesicles are in a crowded environment. In support of this result, it should be noted that Lu *et al.* had found volume fractions of vesicles in the order of 60% for Infasurf, another surfactant substitute [33] and more recently Ciutara and Zasadzinski reported values in the range 40-50% for Curosurf® and Survanta [34], in good agreement with our results. The $\phi(c)$-relationship will be further useful to express the measured viscosity as a function of volume fraction and to compare it to rheological models of colloids.

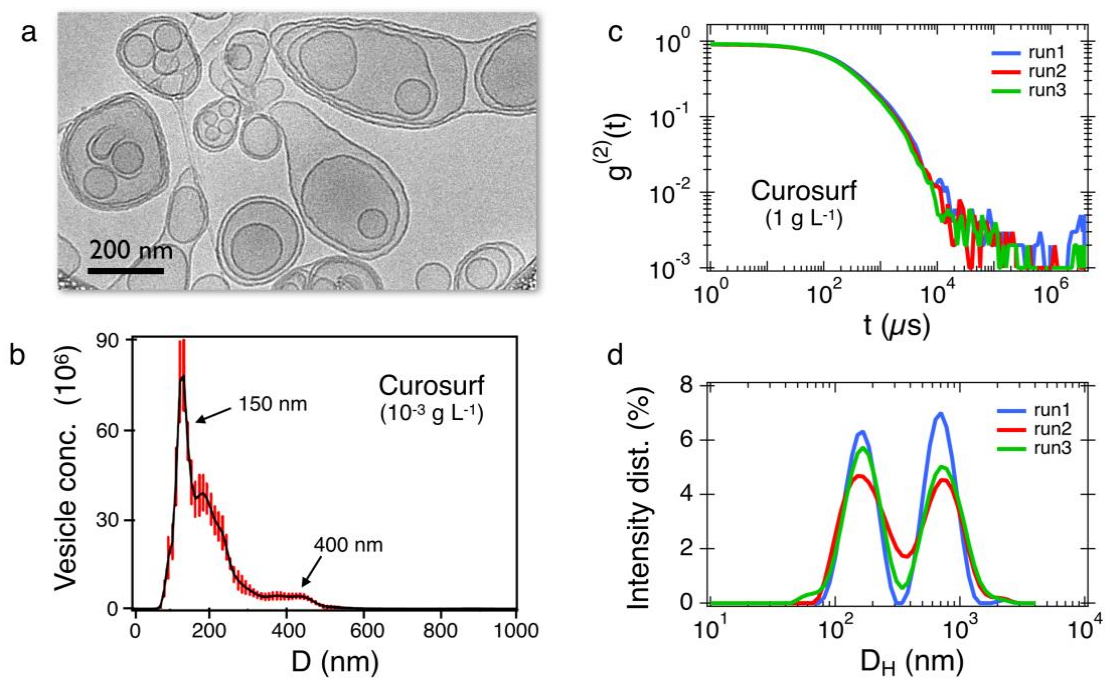

***Figure 2 : a)*** *Cryo-TEM image of Curosurf® at the concentration of 5 g L$^{-1}$.* ***b)*** *Vesicle number distribution determined by nanoparticle tracking analysis (NTA) on Curosurf® at 10$^{-3}$ g L$^{-1}$. The distribution (continuous line in black) has been averaged on 4 runs.* ***c)*** *Autocorrelation function $g^{(2)}(t)$ of the scattered light obtained from Curosurf® at 1 g L$^{-1}$.* ***d)*** *Intensity distributions corresponding to the correlograms in* ***c)***. *All experiments were made at room temperature (T = 25 °C).*

**Figs. 2c** and **2d** show the light scattering data in the form of the second order autocorrelation function $g^{(2)}(t)$ and the intensity distribution of the hydrodynamic diameters $D_H$. We also find here data similar to the two previous techniques, with maxima in intensity at 170 and 700 nm. As light scattering is sensitive to large objects, this technique highlights the population of vesicles that is larger than the number distributions obtained from Cryo-TEM and NTA. Taken together, the data on Curosurf® structure reveals an assembly of lipids in lamellar bilayers, and a wide distribution of vesicle sizes from 50 nm and a few micrometers, with the smallest vesicles being the most numerous.

**III.2 – Viscosity of Curosurf® dispersion as a function of the concentration**

During breathing, the volume of the alveolus (**Fig. 1**) varies by about 10% [35], leading to the recirculation and flow of pulmonary surfactant at the interface with air. Such a process is indeed dependent on the surface tension and on the viscosity of the fluid. Surprisingly, there are currently few rheology measurements of endo- or exogenous pulmonary surfactant [26, 33, 34]. As a vesicle dispersion, pulmonary surfactant is expected to behave as a viscous liquid under physiological conditions, i.e. at a lipid/protein concentration of 40 g L$^{-1}$ [4-7]. Given the cost of Curosurf® samples, it is more convenient to use microrheology rather than conventional rheometry for a systematic study of the viscosity. To this aim we applied the technique of magnetic rotational spectrometry (MRS) in which micro-sized magnetic wires are subjected to a rotating magnetic field at increasing angular frequency $\omega$. The wires are 10-100 µm long, with a diameter of about 1 µm. Due to their superparamagnetic properties, these objects can be actuated, leading to a propeller-like motion. The generic model of magnetic wire embedded in a viscous Newton liquid of static shear viscosity $\eta$ has been derived [36, 37]. It predicts that at low frequency, the wire rotates in phase with the field; the motion is synchronous. With increasing $\omega$, the friction torque increases and above a critical value noted $\omega_C$ the wire undergoes a transition towards an asynchronous regime at [36, 37]:

$$\omega_C = \frac{3}{8\mu_0} \frac{\Delta\chi}{\eta} \frac{B^2}{L^{*2}} \qquad (1)$$

In Eq. 1, $\mu_0$ is the vacuum permeability, $\Delta\chi$ the anisotropy of susceptibility between parallel and perpendicular wire direction and $L^* = L/D\sqrt{g(L/D)}$, with $g(x) = ln(x) - 0.662 + 0.917x - 0.050x^2$ [27]. With MRS, the simplest method to determine $\eta$ is to measure the critical frequency $\omega_C$ as a function of the reduced wire length $L^*$, and to verify the $L^{*-2}$-dependence in Eq. 1. The static viscosity is then retrieved from the value of the prefactor featuring in Eq. 1, $3\Delta\chi B^2/8\mu_0\eta$.

MRS microrheology experiments have been performed at 25 °C as a function of the lipid/protein concentration, from 0 to 80 g L$^{-1}$. In this range, experiments have displayed a wire behavior that is characteristic of Newtonian fluids. At the highest concentrations (80 g L$^{-1}$) the viscosity was found to be 22 ± 3 mPa s, around 25 times that of the solvent. **Fig. 3** shows the variation of the viscosity of Curosurf® dispersions as a function of the vesicle volume fraction $\phi$, this later quantity being computed from the NTA results. At $\phi$ less than 10%, the data show Einstein-like linear behavior characteristic of dilute colloids. Above, the increase is stronger and attributed to interparticle interaction [38]. In parallel we performed macroscopic rheometry measurements with a Physica RheoCompass MCR 302 (Anton Paar) working with a 50 mm wide cone-and-plate geometry at the two reference concentrations, 40 and 80 g L$^{-1}$. Both frequency and shear rate sweeps were performed at 25 and 37 °C, leading to data of the loss modulus $G''(\omega)$ and shear stress $\sigma(\dot{\gamma})$, where $\dot{\gamma}$ denotes the shear rate.

The viscosity of Curosurf® did not show much difference between room and physiological temperature. These results disclose a Newtonian behavior for the 40 g L$^{-1}$ sample and a slightly viscoelastic behavior for the 80 g L$^{-1}$. In contrast to Ciutara and Zasadzinski [34], no yield stress behavior was found in the Curosurf® samples studied here. Of note here, the data from cone-and-plate rheology show good agreement with those of MRS technique (**Fig. 3b**). We have reported on the figure the Krieger-Dougherty law found for a wide variety of colloidal particles [39]:

$$\eta(\phi) = \eta_S \left(1 - \frac{\phi}{\phi_m}\right)^{-2} \qquad (2)$$

In Eq. 2, $\eta_S$ is the solvent viscosity and $\phi_m$ is the maximum-packing volume fraction at the sol-gel transition. Dispersions above $\phi_m$ exhibit the properties of yield stress materials, *i.e.* materials that can flow if they are mechanically subjected to shear stresses above a critical value. For Curosurf®, we found a maximum-packing volume fraction of $\phi_m$ = 0.62, in good agreement with experiments on concentrated colloidal dispersion and theory [40, 41].

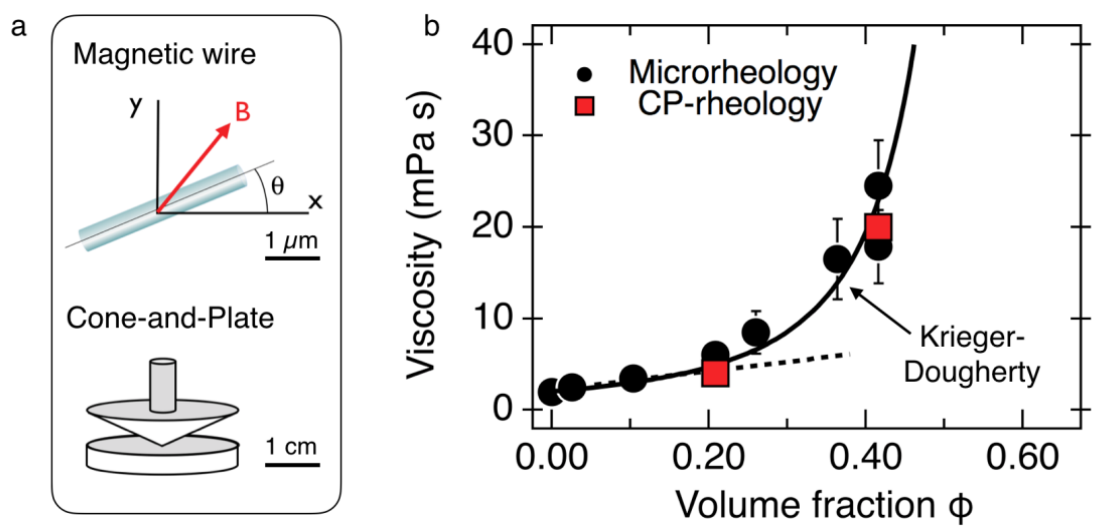

**Figure 3 :** *a)* Tools used to measure the static viscosity of Curosurf® dispersions. Microrheology experiments on magnetic wire can be performed with samples of less than 10 µL, whereas the Cone-and-Plate (CP) rheometer require around 1 mL of sample. *b)* Static viscosity of Curosurf® dispersions as a function of the volume fraction in vesicles. The dotted line is obtained from the Einstein linear dependence and the continuous line is from the Krieger-Dougherty law (Eq. 2) [38, 39].

**III.3 – Impact of alumina nanoplatelets on the surfactant rheology**
*III.3.1 – Modification of Curosurf® microstructrure in presence of alumina nanoplatelets*
Here, we estimate the concentrations of nanoparticles in the alveolar region of the lungs that are consistent with actual air pollution by PM$_{10}$. For this purpose, we consider the case of airborne particle densities in the 100 µg m$^{-3}$ range, which are values commonly found, and in some cases exceeded in polluted city centers [42, 43]. According to European air quality indices, such PM$_{10}$ densities correspond, to a very poor to hazardous air quality for humans. Based on estimations made by Schleh *et al.* [30] on sub-100 nm TiO$_2$ nanoparticles, the corresponding mass of particles deposited in the alveoli amounts at 360 µg per day. Similar calculations by Geiser and Kreyling [10] using a particle deposition fraction of 0.30 lead to a daily dose of 280 µg deposited in the alveolar region. Assuming that the volume of pulmonary surfactant in humans is around 25 mL and distributed over 300 million alveoli [2], we end up with a pulmonary surfactant volume of 80 pL per alveolus, and a daily concentration in the pulmonary surfactant of $(12 - 16) \times 10^{-3}$ g L$^{-1}$. If, in addition, one assumes that these daily levels can be cumulative, it is reasonable to consider as realistic a broader range for the NP concentration, say between $c_{NP} = 10^{-3}$ and 0.1 g L$^{-1}$. In this type of estimation, it should be kept in mind that previous estimates depend on a variety of factors, such as the nature and physico-chemical properties of the particles and that some of these factors are still under debate [10].

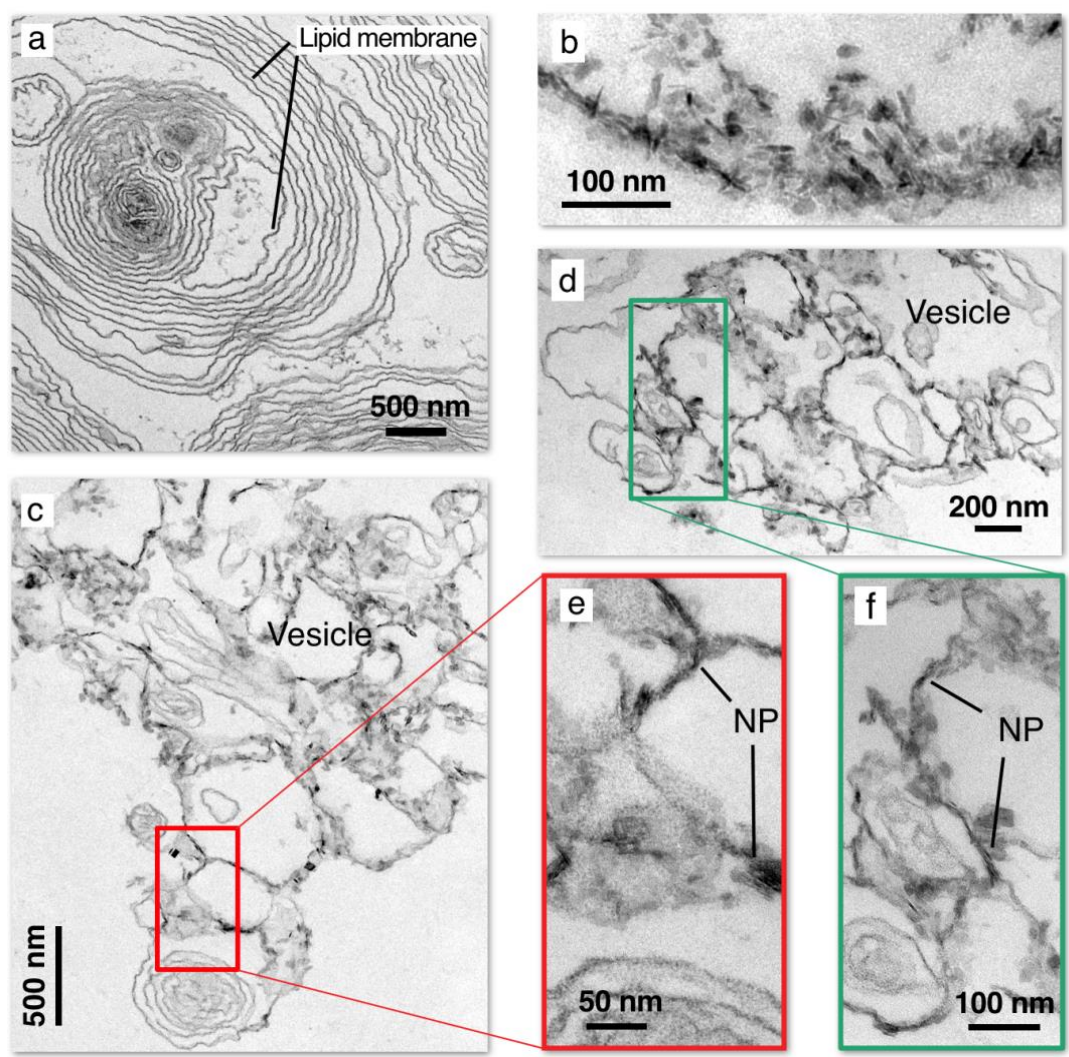

***Figure 4 : a)*** *Transmission electron microscopy (TEM) images of a Curosurf® dispersion at lipid/protein concentration of 80 g L$^{-1}$ showing multilamellar and multivesicular vesicles.* ***b)*** *TEM image of positively charged alumina nanoplatelets of size 40 nm (platelet diameter) and dispersity s = 0.30. s is defined as the ratio between the standard deviation and the mean.* ***c,d)*** *TEM images of a multivesicular vesicle after interacting with Al$_2$O$_3$-NPs. The lipid/protein and alumina concentrations are 80 g L$^{-1}$ and $c_{NP} = 1$ g L$^{-1}$, respectively.* ***e,f)*** *Boxed areas showing a close-up view of the multivesicular vesicles in* ***c)*** *and* ***d)*** *respectively showing Al$_2$O$_3$-NPs adsorbed at the lipid membranes.*

To evaluate the impact of $Al_2O_3$-NPs on Curosurf® we performed TEM and rheology experiments on pristine Curosurf® and on $Al_2O_3$/Curosurf® mixtures. **Fig. 4a** shows a TEM image of a Curosurf® dispersion at a lipid/protein concentration of 80 g L$^{-1}$ after fixation with glutaraldehyde and labeling with osmium. Following fixation, the sample was microtomized into 70 nm sections, which were then deposited on a TEM grid. The figure shows the presence of a 2.7 µm large multivesicular vesicle together with several enclosed uni/multilamellar substructures. Lipid membranes are clearly visible, as are neighboring vesicles, the proximity of which confirms the finding of a crowded state for the clinical formulation. The observed fluctuations in the membrane contours are related to surface tension or elastic properties of the lipid bilayer. Alumina NPs obtained by TEM from a $c_{NP}$ = 1 g L$^{-1}$ dispersion deposited on a TEM grid show anisotropic nanoparticles, with a main diameter of 40 nm, and dispersity 0.30 (**Fig. 4b**). The thickness of these nanoplatelets is estimated to be about 10 nm [13]. **Fig. 4c** and **4d** displays a TEM image of a multivesicular vesicle after interacting with $Al_2O_3$-NPs. Compared to **Fig. 4a**, the structure has been profoundly modified. The image reveals a froth-like arrangement with membrane compartments that no longer resemble those of pristine vesicles. On the other hand, the contrast of the membranes is more irregular, a phenomenon that is linked to the presence of nanoplatelets adsorbed on the lipid bilayers. Enlarged views of selected areas displaying $Al_2O_3$-NPs are shown in **Fig. 4e** and **4f**. In these examples, it is found that the NPs come into contact and adhere to the lipid membranes, a result that has been attributed to electrostatic interaction [13]. We recall here that the surface charge density of alumina NPs and Curosurf® are of +7.3 *e* nm$^{-2}$ and -0.19 *e* nm$^{-2}$ respectively [22], and we ascribe the patterned structure of **Fig. 4e** and **4f** to the high density of positive charges carried by the alumina NPs.

*III.3.2 – Rheology of $Al_2O_3$/Curosurf®*
It was by observing $Al_2O_3$/Curosurf® samples under optical microscopy that we noticed an abnormal behavior regarding the vesicle Brownian dynamics. The upper panel of **Fig. 5a** shows a 40× magnified image of a 40 g L$^{-1}$ dispersion obtained by phase contrast microscopy. At this magnification, we observe a continuum of intensity that is spatially heterogeneous, and that comes from vesicles above and below the focal plane of observation. Vesicles in the focal plane are also clearly identifiable. In this configuration we collected the light transmitted by the camera over 5 minutes. This intensity in **Fig. 5a** (lower panel) exhibits large fluctuations coming from the Brownian motion of the vesicular objects. For Curosurf® the amplitude distribution of these fluctuations was found to be Gaussian-like [19]. **Fig. 5b** illustrates a zone of identical size for the $Al_2O_3$/Curosurf® mixture at concentration 0.5 g L$^{-1}$ / 40 g L$^{-1}$. The samples with and without NPs appear identical under phase contrast microscopy. However, strong differences are observed in the fluctuations of the transmitted light (lower panel of **Fig. 5b**). The intensity collected for $Al_2O_3$/Curosurf® reveals fluctuation which amplitudes are barely perceptible to the eye. These results suggest that the addition of NPs to Curosurf® has induced a transition to a dynamic arrested state, which we now investigate in microrheology.

In the high frequency range, the generic behavior of magnetic wires subjected to a rotating magnetic field can be summarized as follows [19]. Purely viscous fluids undergo a transition towards asynchronous motion with back-and-forth oscillations above a critical frequency $\omega_C$, whereby the oscillation amplitude decreases as $\theta_B(\omega) \sim \omega^{-1}$. For viscoelastic or soft solids, an asynchronous regime is found at all frequency and $\theta_B(\omega)$ reflects the elastic response of the material. **Fig. 5c** displays the wire oscillation amplitudes as a function of the reduced frequency $\omega/\omega_C$ for Curosurf® 40 and 80 g L$^{-1}$. Wires between 20 and 80 µm were used in these measurements. At 40 g L$^{-1}$, $\theta_B(\omega)$ decreases with increasing frequency in accordance with the Newton constitutive equation prediction [36, 37]. In the $\theta_B(\omega/\omega_C)$-representation used here, the continuous line is obtained with no adjustable parameter and it accounts well for the data. At 80 g L$^{-1}$, a strong decay of $\theta_B(\omega)$ is also observed as a function of frequency. However, it is found to deviate slightly from Newton's prediction, a result that is interpreted as the onset of viscoelasticity. The data is however very close to that of the 40 g L$^{-1}$ sample, which allows us to conclude that Curosurf® behaves globally as a purely viscous fluid in the studied concentration range. When minute amounts of Al$_2$O$_3$-NPs are added, typically for concentrations below 0.1 g L$^{-1}$, the same Newtonian fluid-like behavior was also observed [19].

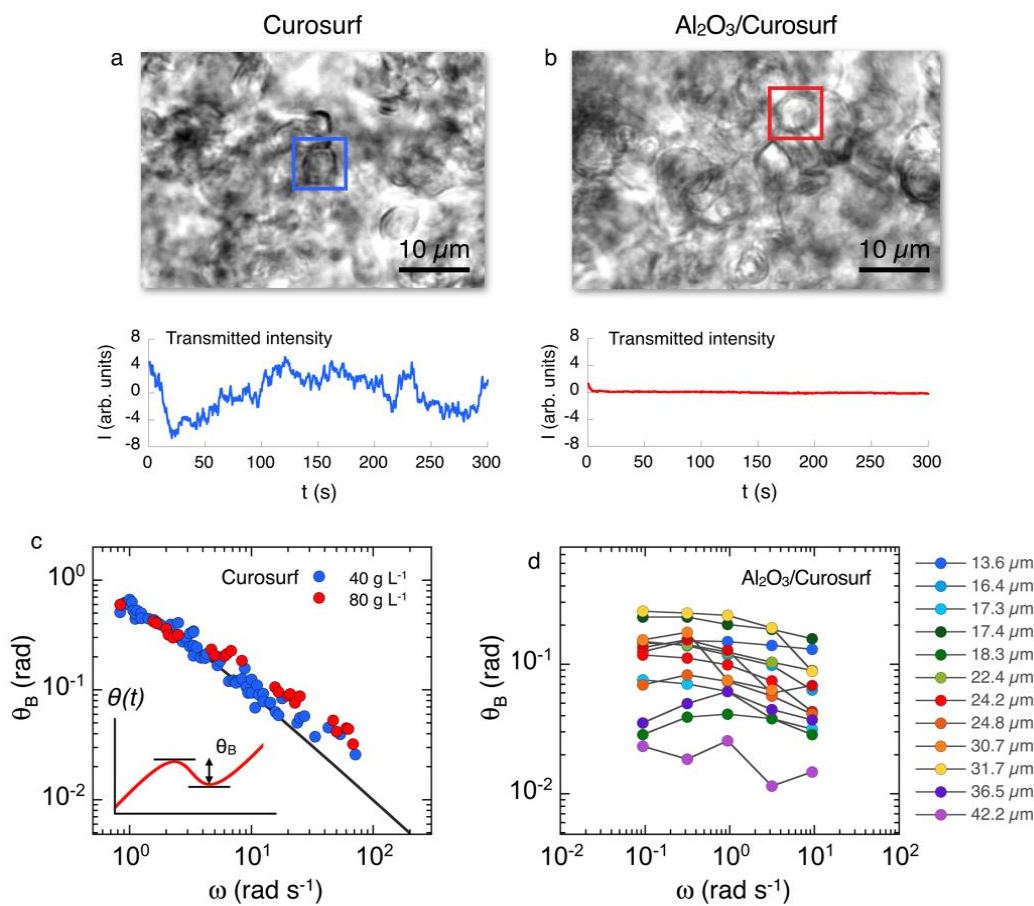

*Figure 5 : a)* Upper panel: Phase contrast microscopy images of a 40 g L-1 Curosurf® dispersion. The square indicates the domain where transmitted light intensity (lower panel) was collected as a function of time. *b)* Same as in *a)* for Curosurf® loaded with 0.5 g L$^{-1}$ of alumina nanoplatelets. *c)* Oscillation amplitude $\theta_B(\omega/\omega_C)$ observed in the asynchronous regime for Curosurf® at 40 and 80 g L$^{-1}$ (T = 25 °C). The continuous line is the constitutive equation solution for Newtonian fluid. Inset: Time dependence of the wire orientation angle $\theta(t)$ showing the definition of the back angle $\theta_B$. *d)* Oscillation amplitude $\theta_B(\omega)$ measured in Curosurf® loaded with alumina particles ( $c_{NP}$ = 0.19 g L$^{-1}$) using 12 wires of lengths between 13.6 and 42.2 µm.

For concentrations above $c_{NP}$ = 0.1 g L$^{-1}$ however, the dispersions exhibit the arrested dynamic regime described in **Fig. 5b**, and the wire motion has the signature of a soft solid material. **Fig. 5d** displays the oscillation amplitude $\theta_B(\omega)$ as a function of the frequency at $c_{NP}$ = 0.49 g L$^{-1}$ and for wires comprised between 13 and 42 µm. It is found that $\theta_B(\omega)$ exhibits a plateau at low frequency, noted $\theta_{B,eq}$ followed by a modest decrease above 1 rad s$^{-1}$. According to the constitutive equations developed for soft solids [29], the low frequency limit is linked to the equilibrium storage modulus $G_{eq}$ through the relation:

$$\lim_{\omega \to 0} \theta_B(\omega) = \theta_{B,eq} = 3\mu_0 \Delta\chi H^2 / 4L^{*2} G_{eq} \qquad (Eq.\,3)$$

For a soft solid, $G_{eq}$ characterizes the quasi-static elastic response of the material, *i.e.* as $\omega$ goes to zero. Similarly, the infinite frequency limit of the oscillation amplitude is related to the instantaneous elastic shear modulus $G$ by the expression:

$$\lim_{\omega \to \infty} \theta_B(\omega) = \frac{\theta_0 \theta_{eq}}{\theta_0 + \theta_{eq}} \text{ and } \theta_0 = 3\Delta\chi B^2 / 4\mu_0 G L^{*2} \qquad (4)$$

The analysis of the behavior of about ten wires in the samples at $c_{NP}$ = 0.19 and 0.49 g L$^{-1}$ leads to elastic modulus values in the range 0.02 – 0.2 Pa. These elasticity values are low, and the dispersions are therefore considered as soft solids. **Table 1** compares the microrheology results obtained on Curosurf® samples without and with Al$_2$O$_3$-NPs. We attribute the sol-gel transition to the formation of a cross-linked network between the Curosurf® vesicles, the cross-linkers being the alumina NPs, as suggested by the results in **Figs. 4c and 4d**. Under such conditions, it is conceivable that the presence of large numbers of NPs in the most distal region of the lungs could have harmful effects on the recirculation of pulmonary surfactant occurring in the lumen of the alveoli.

| Sample | Model fluid | $\eta$ (mPa s) | Sync./Async. transition | $G_{eq}$ (Pa) | $G$ (Pa) |
|---|---|---|---|---|---|
| Curosurf® 40 g L$^{-1}$ | Newton | 5 ± 1 | yes | 0 | 0 |
| Curosurf® 80 g L$^{-1}$ | Newton | 22 ± 3 | yes | 0 | 0 |
| Al$_2$O$_3$/Curosurf® 0.19/44 g L$^{-1}$ | Soft solid | ∞ | no | 0.016 ± 0.01 | 0.11 ± 0.03 |
| Al$_2$O$_3$/Curosurf® 0.49/44 g L$^{-1}$ | Soft solid | ∞ | no | 0.18 ± 0.05 | 0.24 ± 0.06 |

**Table 1:** *Summary of microrheology experiments conducted on Curosurf® at physiological concentration (40 g L$^{-1}$) and on the clinical formulation (80 g L$^{-1}$) without added particles, as well as on 44 g L$^{-1}$ formulations containing $c_{NP}$ = 0. 19 and 0.49 g L$^{-1}$ of alumina nanoplatelets.*

## IV – Conclusion

We report on the structural and dynamical properties of the pulmonary surfactant substitute Curosurf® which is administered to premature babies as prophylaxis for the treatment of respiratory distress syndrome at birth [14]. Using spectrometry for deciphering Curosurf® structure, we find that the clinical dispersion is made of phospholipids assembled in multivesicular vesicles characterized by a wide distribution of sizes between 50 nm and 5 µm. Using an active microrheology technique refereed to magnetic rotational spectrometry and based on the magnetic rotation of micron-sized wires, we measure the Curosurf® viscosity for concentration ranging from 0 to 80 g L$^{-1}$ and found a Newtonian behavior, the static shear viscosity increasing according to the Krieger-Dougherty behavior. Upon addition of 40 nm alumina nanoplatelets to Curosurf® at concentration above 0.1 g L$^{-1}$, we observe a significant change in the structure and rheology of the dispersion. The change in the flow properties is consistent with the appearance of a gel phase, with an infinite viscosity and a zero-frequency elastic modulus. The MRS technique is quantitative and allows to determine the elastic moduli, around 0.1 Pa. This outcome is interpreted by the interaction of the $Al_2O_3$-NPs that establish a crosslink between the lipid vesicles, forming thereby a percolating network. Under such conditions, it is conceivable that the presence of large numbers of NPs in the distal region of the lungs could have harmful effects on the recirculation of pulmonary surfactant occurring in the lumen of the alveoli. This result could indicate a potential toxicity associated with the modification of the lung fluid rheology.

## Acknowledgments

We thank Mostafa Mokhtari, Jesus Perez-Gil, Chloé Puisney, Milad Radiom and Nicolas Tsapis for fruitful discussions. Imane Boucema is acknowledged for letting us use the Anton Paar rheometer for the cone-and-plate rheology. ANR (Agence Nationale de la Recherche) and CGI (Commissariat à l'Investissement d'Avenir) are gratefully acknowledged for their financial support of this work through Labex SEAM (Science and Engineering for Advanced Materials and devices) ANR 11 LABX 086, ANR 11 IDEX 05 02. We acknowledge the ImagoSeine facility (Jacques Monod Institute, Paris, France), and the France BioImaging infrastructure supported by the French National Research Agency (ANR-10-INSB-04, « Investments for the future »). This research was supported in part by the Agence Nationale de la Recherche under the contract ANR-13-BS08-0015 (PANORAMA), ANR-12-CHEX-0011 (PULMONANO), ANR-15-CE18-0024-01 (ICONS), ANR-17-CE09-0017 (AlveolusMimics) and by Solvay.

## TOC Image

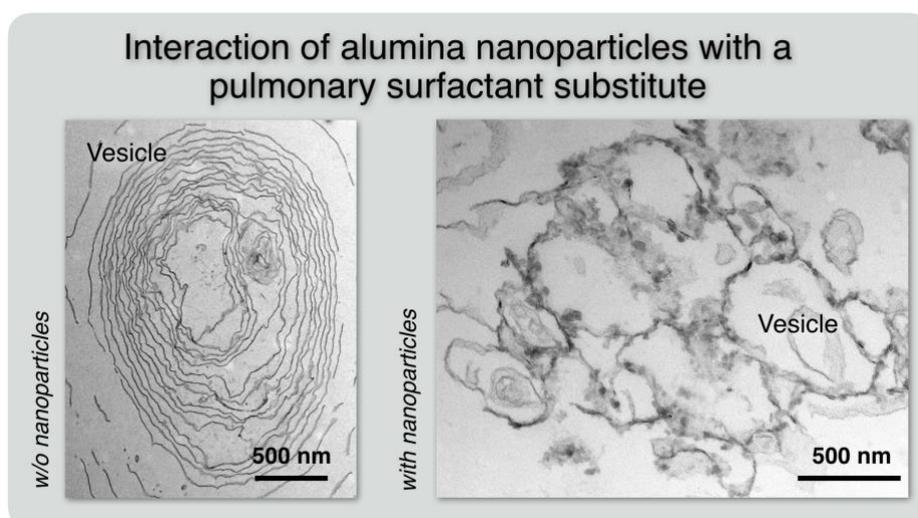

# References

[1] P. Bajaj, J.F. Harris, J.H. Huang, P. Nath, R. Iyer, Advances and Challenges in Recapitulating Human Pulmonary Systems: At the Cusp of Biology and Materials, ACS Biomater. Sci. Eng, 2 (2016) 473-488.

[2] R.H. Notter, Lung Surfactant: Basic Science and Clinical Applications, CRC Press, Boca Raton, FL, 2000.

[3] J.C. Castillo-Sánchez, N. Roldán, B. García-Álvarez, E. Batllori, A. Galindo, A. Cruz, J. Pérez-Gil, The highly packed and dehydrated structure of preformed unexposed human pulmonary surfactant isolated from amniotic fluid, Am. J. Physiol.: Lung Cell. Mol. Physiol., 322 (2022) L191-l203.

[4] N. Hobi, G. Siber, V. Bouzas, A. Ravasio, J. Perez-Gil, T. Haller, Physiological Variables Affecting Surface Film Formation by Native Lamellar Body-Like Pulmonary Surfactant Particles, Biochim. Biophys. Acta, Biomembr., 1838 (2014) 1842-1850.

[5] T. Kobayashi, A. Shido, K. Nitta, S. Inui, M. Ganzuka, B. Robertson, The critical concentration of surfactant in fetal lung liquid at birth, Resp. Physiol., 80 (1990) 181-192.

[6] J.F. Lewis, A.H. Jobe, Surfactant and the Adult Respiratory-Distress Syndrome, Am. Rev. Respir. Dis., 147 (1993) 218-233.

[7] M. Numata, P. Kandasamy, D.R. Voelker, Anionic Pulmonary Surfactant Lipid Regulation of Innate Immunity, Expert Rev. Respir. Med., 6 (2012) 243-246.

[8] E. Lopez-Rodriguez, J. Perez-Gil, Structure-Function Relationships in Pulmonary Surfactant Membranes: From Biophysics to Therapy, Biochim. Biophys. Acta, Biomembr., 1838 (2014) 1568-1585.

[9] A.R. Associates, Multi-Path Particle Deposition model (MPPD), 2009.

[10] M. Geiser, W.G. Kreyling, Deposition and biokinetics of inhaled nanoparticles, Part. Fibre Toxicol., 7 (2010).

[11] A. Hidalgo, A. Cruz, J. Perez-Gil, Pulmonary Surfactant and Nanocarriers: Toxicity *versus* Combined Nanomedical Applications, Biochim. Biophys. Acta, Biomembr., 1859 (2017) 1740-1748.

[12] F. Mousseau, J.F. Berret, The Role of Surface Charge in the Interaction of Nanoparticles with Model Pulmonary Surfactants, Soft Matter, 14 (2018) 5764-5774.

[13] F. Mousseau, R. Le Borgne, E. Seyrek, J.-F. Berret, Biophysicochemical Interaction of a Clinical Pulmonary Surfactant with Nanoalumina, Langmuir, 31 (2015) 7346-7354.

[14] J. Egberts, J.P. de Winter, G. Sedin, M.J. de Kleine, U. Broberger, F. van Bel, T. Curstedt, B. Robertson, Comparison of Prophylaxis and Rescue Treatment with Curosurf in Neonates less than 30 Weeks' Gestation: A Randomized Trial, Pediatrics, 92 (1993) 768-774.

[15] T. Curstedt, H.L. Halliday, C.P. Speer, A Unique Story in Neonatal Research: The Development of a Porcine Surfactant, Neonatology, 107 (2015) 321-329.

[16] I. Panaiotov, T. Ivanova, J. Proust, F. Boury, B. Denizot, K. Keough, S. Taneva, Effect of Hydrophobic Protein Sp-C on Structure and Dilatational Properties of The Model Monolayers of Pulmonary Surfactant, Colloids Surf. B, 6 (1996) 243-260.

[17] S. Sweeney, B.F. Leo, S. Chen, N. Abraham-Thomas, A.J. Thorley, A. Gow, S. Schwander, J.F.J. Zhang, M.S.P. Shaffer, K.F. Chung, M.P. Ryan, A.E. Porter, T.D. Tetley, Pulmonary surfactant mitigates silver nanoparticle toxicity in human alveolar type-I-like epithelial cells, Colloids Surf. B, 145 (2016) 167-175.

[18] W. Wohlleben, M.D. Driessen, S. Raesch, U.F. Schaefer, C. Schulze, B. von Vacano, A. Vennemann, M. Wiemann, C.A. Ruge, H. Platsch, S. Mues, R. Ossig, J.M. Tomm, J. Schnekenburger, T.A.J. Kuhlbusch, A. Luch, C.M. Lehr, A. Haase, Influence of agglomeration and specific lung lining lipid/protein interaction on short-term inhalation toxicity, Nanotoxicology, 10 (2016) 970-980.

[19] L.-P.-A. Thai, F. Mousseau, E. Oikonomou, M. Radiom, J.-F. Berret, Effect of Nanoparticles on the Bulk Shear Viscosity of a Lung Surfactant Fluid, ACS Nano, 14 (2020) 466-475.

[20] L. Camacho, A. Cruz, R. Castro, C. Casals, J. Pérez-Gil, Effect of Ph on the Interfacial Adsorption Activity of Pulmonary Surfactant, Colloids Surf. B, 5 (1996) 271-277.

[21] F. Mousseau, L. Vitorazi, L. Herrmann, S. Mornet, J.F. Berret, Polyelectrolyte Assisted Charge Titration Spectrometry: Applications to Latex and Oxide Nanoparticles, J. Colloid Interface Sci., 475 (2016) 36-45.

[22] F. Mousseau, E.K. Oikonomou, V. Baldim, S. Mornet, J.F. Berret, Nanoparticle-Lipid Interaction: Job Scattering Plots to Differentiate Vesicle Aggregation from Supported Lipid Bilayer Formation, Colloids Interfaces, 2 (2018).
[23] J. Dubochet, On the Development of Electron Cryo-Microscopy (Nobel Lecture), Angew. Chem. Int. Ed., 57 (2018) 10842-10846.
[24] E.K. Oikonomou, K. Golemanov, P.-E. Dufils, J. Wilson, R. Ahuja, L. Heux, J.-F. Berret, Cellulose Nanocrystals Mimicking Micron-Sized Fibers to Assess the Deposition of Latex Particles on Cotton, ACS Appl. Polym. Mat., 3 (2021) 3009-3018.
[25] L.P.A. Thai, F. Mousseau, E.K. Oikonomou, J.F. Berret, On the Rheology of Pulmonary Surfactant: Effects of Concentration and Consequences for the Surfactant Replacement Therapy, Colloids and Surfaces B-Biointerfaces, 178 (2019) 337-345.
[26] D.M. King, Z.D. Wang, H.J. Palmer, B.A. Holm, R.H. Notter, Bulk Shear Viscosities of Endogenous and Exogenous Lung Surfactants, Am. J. Physiol.: Lung Cell. Mol. Physiol., 282 (2002) L277-L284.
[27] L. Chevry, N.K. Sampathkumar, A. Cebers, J.F. Berret, Magnetic Wire-Based Sensors for the Microrheology of Complex Fluids, Phys. Rev. E, 88 (2013) 062306.
[28] J.-F. Berret, Local Viscoelasticity of Living Cells Measured by Rotational Magnetic Spectroscopy, Nat. Commun., 7 (2016) 10134.
[29] F. Loosli, M. Najm, R. Chan, E. Oikonomou, A. Grados, M. Receveur, J.-F. Berret, Wire-Active Microrheology to Differentiate Viscoelastic Liquids from Soft Solids, ChemPhysChem, 17 (2016) 4134-4143.
[30] C. Schleh, C. Muhlfeld, K. Pulskamp, A. Schmiedl, M. Nassimi, H.D. Lauenstein, A. Braun, N. Krug, V.J. Erpenbeck, J.M. Hohlfeld, The Effect of Titanium Dioxide Nanoparticles on Pulmonary Surfactant Function and Ultrastructure, Respir. Res., 10 (2009) 90.
[31] D. Waisman, D. Danino, Z. Weintraub, J. Schmidt, Y. Talmon, Nanostructure of the Aqueous Form of Lung Surfactant of Different Species Visualized by Cryo-Transmission Electron Microscopy, Clin. Physiol. Funct. Imaging, 27 (2007) 375-380.
[32] J. Bernardino de la Serna, R. Vargas, V. Picardi, A. Cruz, R. Arranz, J.M. Valpuesta, L. Mateu, J. Perez-Gil, Segregated ordered lipid phases and protein-promoted membrane cohesivity are required for pulmonary surfactant films to stabilize and protect the respiratory surface, Faraday Discuss., 161 (2013) 535-548.
[33] K.W. Lu, J. Perez-Gil, H.W. Taeusch, Kinematic Viscosity of Therapeutic Pulmonary Surfactants with Added Polymers, Biochim. Biophys. Acta, Biomembr., 1788 (2009) 632-637.
[34] C.O. Ciutara, J.A. Zasadzinski, Bilayer aggregate microstructure determines viscoelasticity of lung surfactant suspensions, Soft Matter, 17 (2021) 5170-5182.
[35] S. Chang, N. Kwon, J. Kim, Y. Kohmura, T. Ishikawa, C.K. Rhee, J.H. Je, A. Tsuda, Synchrotron x-ray imaging of pulmonary alveoli in respiration in live intact mice, Sci. Rep., 5 (2015) 8760.
[36] B. Frka-Petesic, K. Erglis, J.-F. Berret, A. Cebers, V. Dupuis, J. Fresnais, O. Sandre, R. Perzynski, Dynamics of paramagnetic nanostructured rods under rotating field, J. Magn. Magn. Mater., 323 (2011) 1309-1313.
[37] G. Helgesen, P. Pieranski, A.T. Skjeltorp, Nonlinear Phenomena in Systems of Magnetic Holes, Phys. Rev. Lett., 64 (1990) 1425-1428.
[38] H.M. Shewan, J.R. Stokes, Analytically predicting the viscosity of hard sphere suspensions from the particle size distribution, J. Non-Newtonian Fluid Mech., 222 (2015) 72-81.
[39] I.M. Krieger, T.J. Dougherty, A Mechanism for Non-Newtonian Flow in Suspensions of Rigid Spheres, Trans. Soc. Rheol., 3 (1959) 137-152.
[40] K.W. Desmond, E.R. Weeks, Influence of particle size distribution on random close packing of spheres, Phys. Rev. E, 90 (2014) 022204.
[41] S.E. Phan, W.B. Russel, J.X. Zhu, P.M. Chaikin, Effects of polydispersity on hard sphere crystals, J. Chem. Phys., 108 (1998) 9789-9795.
[42] J. Lelieveld, J.S. Evans, M. Fnais, D. Giannadaki, A. Pozzer, The Contribution of Outdoor Air Pollution Sources to Premature Mortality on a Global Scale, Nature, 525 (2015) 367-+.

[43] A. van Donkelaar, R.V. Martin, M. Brauer, R. Kahn, R. Levy, C. Verduzco, P.J. Villeneuve, Global estimates of ambient fine particulate matter concentrations from satellite-based aerosol optical depth: development and application, Environ. Health Perspect., 118 (2010) 847-855.